\documentclass[amsmath,amssymb,nofootinbib,superscriptaddress,showkeys]{article}
\usepackage{pslatex}
\usepackage{palatino}
\usepackage{mathtools}
\def\bea{\begin{eqnarray}}
\def\eea{\end{eqnarray}}

\def\slashchar#1{\setbox0=\hbox{$#1$}
    \dimen0=\wd0 \setbox1=\hbox{/} \dimen1=\wd1
   \ifdim\dimen0>\dimen1 \rlap{\hbox to \dimen0{\hfil/\hfil}} #1
   \else  \rlap{\hbox to \dimen1{\hfil$#1$\hfil}} / \fi}

\begin{document}

\title{\bf  \sl Some Three-body force cancellations in Chiral Lagrangians~
\footnote{Dedicated to the memory of Juan Antonio Morente Chiquero
  (1955-2012).}}
\author{Enrique  Ruiz
  Arriola \\
\normalsize Departamento de F\'isica At\'omica, Molecular y Nuclear \\ \normalsize and Instituto Carlos I de F\'isica Te\'orica y Computacional \\
\normalsize Universidad de Granada,
  E-18071 Granada, Spain.}
%
\maketitle

\begin{abstract} 
The cancellation between off-shell two body forces and three body
forces implies a tremendous simplification in the study of three body
resonances in two meson-one baryon systems.  While this can be done by
means of Faddeev equations we provide an alternative and simpler
derivation using just the chiral Lagrangean and the field
reparameterization invariance.
\end{abstract}


\section{Introduction}

The main and stunning difference between point particles and waves is
that the latter contain an effective size corresponding to their
wavelength. This provides a natural resolution scale, $\lambda$, {\it
  below} which details cannot be resolved. For instance, the
boundaries of macroscopic matter objects do not show up their
roughness below the visible light wavelength of the order of
$\lambda_\gamma \sim 400-800 nm$ and hence appear as locally
smooth. In non-relativistic Quantum Mechanics, due to the
wave-particle duality, the de Broglie wavelengh sets up the scale
$\lambda_{\rm dB} = \hbar / M v $ and becomes large for slow enough
particles. In Relativistic Quantum Field Theory and because of
microcausality, interactions are generated by particle exchange and
the corresponding Compton wavelength $\lambda_C = \hbar / m c$
provides the range of the interaction, exemplified by the Yukawa-like
potential $V(r) \sim e^{-r/\lambda_C}/r $.

Thus, if we think of N interacting composite quantum particles with
rest masses $M_i$ and typical momenta $p_i$ and interacting through
exchanges or particles with masses $m_j$ we have that for $p_{\rm max}
\equiv {\rm max} \, |p_i-p_j| \ll m_{\rm min} \equiv {\rm min} \, m_i
$ we expect not to resolve the precise form of the interaction and the
few-body problem may be analyzed within a systematic expansion in
$p_{\rm max}/m_{\rm min}$. The coefficients of such an expansion are
called low energy constants (LEC's). These LEC's implement shape
independence, short-distance insensitivity and effective elementarity
and summmarize all effects not explicitly taken into account.
Nonetheless, they also depend on the resolution scale $\lambda$ and a
too coarse resolution may not allow to distinguish between N-body
correlations from the difficult N-body forces due to (N-1) irreducible
particle exchange (with range $\sim \lambda_C/N)$. A concise form of
summarising all these features is by using an Effective Field Theory
(EFT)~\cite{Weinberg:1978kz} where the renormalization scale $\mu =
\hbar c / \lambda$ is used instead. A more tangible approach uses the
concept of coarse grained interactions (see
e.g.~\cite{NavarroPerez:2011fm}  for a Nuclear Physics setup).

While this discussion is quite general we will focus here on the
application to hadronic and relativistic systems described by quantum
fields and show how further simplifications can arise in modern chiral
Lagrangians which effectively describe scattering and bound states
involving one baryon and two mesons, such as e.g. the $\pi\pi N$
system. 

\section{Hadronic Interpolating Fields}

Quantum Chromo Dynamics (QCD) is expected to describe all known
composite hadronic systems from the pion to finite nuclei in terms of
$(u,d,s, \dots)$, quarks (or anti-quarks) and gluons.  One can
construct local {\it hadronic} interpolating and composite fields out
of quark fields $q(x)$ and gluon fields $A_\mu(x)$ in terms of
covariant derivatives $D_\mu q = (\partial_\mu +i A_\mu )q$ and field
strength tensor $G_{\mu \nu}= \partial_\mu A_\nu -\partial_\mu A_\nu +
i [A_\mu,A_\nu] $ carrying the same quantum numbers. This of course
generates the problem of operator mixing which resembles the freedom
of choice of a basis in the standard quantum mechanical variational
method. In EFT this is usually reorganized in a dimensional expansion
with growing energy dimensions. For instance, for a scalar-isoscalar
particle with $J^{PC}=0^{++}$ we have the composite field expansion 
in  the resolution wavelength  $\lambda$ 
\begin{eqnarray}
\sigma(x) = Z_2 \lambda^2 \bar q(x) q(x) + Z_3 \lambda^3 G^2 (x) + \lambda^4 Z_4 \bar q (x) D^2 q(x) +
Z_5 \lambda^5 \left[\bar q (x) q(x)\right]^2 + \dots
\end{eqnarray}
where $ G^2(x) = {\rm tr}_c G_{\mu \nu} (x) G^{\mu \nu}(x)$ and
$Z_n(\lambda) $ are dimensionless constants. Physically, the expansion
corresponds to a Fock space decomposition of composite particles which
are treated as elementary with constituents placed at the same point
$x$. The effective elementarity occurs when $ \partial_x \sigma \ll
\sigma/\lambda$. For similar reasons, interactions in an effective
Lagrangian can be written in a dimensional expansion where 1)
classical equations of motion are used and 2) fields may be
reparameterized by any local transformation of the field $\sigma (x)$.
An important issue is that if quantum corrections to the effective
Lagrangian are also suppressed in the resolution scale $\lambda$ a
fully consistent EFT may be built.

However, a direct calculation of Green functions in EFT does not
necessarily guarantee off-shell finiteness from on shell
renormalization conditions (see e.g. Ref.~\cite{Appelquist:1980ae})
and suitable field redefinitions may be requested to ensure off-shell
renormalizability.  The on-shell scheme of Georgi~\cite{Georgi:1991ch}
for EFT's allows to consider on-shell vertices and the problem is
circumvented from the start, since off-shellness cannot be
measured~\cite{Fearing:1999fw} (see also
\cite{Nieves:1999bx}). However, this does not mean that {\it any}
off-shellness can be removed as it was pointed out for chiral two-pion
exchange NN interactions~\cite{Entem:2009mf}.

\section{Chiral Lagrangians}

Chiral Lagrangians in non-linear realizations~\cite{Weinberg:1968de}
(for a review see e.g.~\cite{Pich:1995bw} and references therein)
capture many known relevant features of low energy hadronic physics in
a systematic expansion in $1/f$ ($f \sim 88 {\rm MeV}$ is the pion weak
decay constant for massless quarks). At lowest order it contain
kinetic and mass baryon pieces and meson-baryon interaction terms and
is given by~\cite{Pich:1995bw}
\begin{eqnarray}
{\cal L}_1 = {\rm Tr} \left\{ \bar{B} \left( {\rm i}
\slashchar{\nabla} - M_B \right) B \right\} + 
\frac{1}{2} \, {\cal D} \, {\rm Tr} \left\{ 
\bar{B} \gamma^\mu \gamma_5 \left\{ u_\mu , B
\right\} \right\} + \frac{1}{2} \, {\cal F} \, {\rm Tr} \left\{ 
\bar{B} \gamma^\mu \gamma_5 [ u_\mu ,B] \right\}  \, ,
\label{LB1}
\end{eqnarray}
The meson kinetic and mass pieces and the baryon mass chiral
corrections are second order and read
\begin{eqnarray}
{\cal L}_2 = {f^2 \over 4} {\rm Tr} \left\{ u_\mu^\dagger u^\mu + 
(U^\dagger \chi + \chi^\dagger U ) \right\} -
b_0 {\rm Tr} ( \chi_+ ) {\rm Tr} (\bar B B) - b_1 {\rm Tr} ( \bar B
\chi_+ B ) - b_2 {\rm Tr} ( \bar B B \chi_+ )
\label{LB2}
\end{eqnarray}
where ``Tr'' stands for the trace in $SU(3)$. In addition,
\begin{eqnarray}
\nabla_{\mu} B &=& \partial_{\mu} B + [ \Gamma_\mu,  \, B \, ] \,
,
\quad 
\Gamma_\mu = \frac{1}{2}\, ( \, u^\dagger
\partial_{\mu} u + u \partial_{\mu} u^\dagger  ) \,
,
\nonumber \\
\label{LB1_exp}
U = u^2 &=& e^{ {\rm i} \sqrt{2} \Phi / f } \, , \qquad u_{\mu} = {\rm
i} u ^\dagger \partial_{\mu} U u^\dagger \, \nonumber \\ \chi_+ &=&
u^\dagger \chi u^\dagger + u \chi^\dagger u \, , \qquad \chi = 2 B_0
{\cal M} \, . 
\end{eqnarray} 
$M_B$ is the common mass of the baryon octect, due to spontaneous
chiral symmetry breaking for massless quarks. The $SU(3)$ coupling
constants which are determined by semileptonic decays of hyperons are
${\cal F} \sim 0.46$, ${\cal D} \sim 0.79$ (${\cal F}+{\cal D} = g_{A}
= 1.25$).  The constants $B_0$ and $f$ are not determined by the
symmetry. The current quark mass matrix is ${\cal M}={\rm
  Diag}(m_u,m_d,m_s)$. The parameters $b_0$, $b_1$ and $b_2$ are
coupling constants with dimension of an inverse mass. The values of
$b_1$ and $b_2$ can be determined from baryon mass splittings, whereas
$b_0$ gives an overall contribution to the octect baryon mass
$M_B$\footnote{Using $SU(3)$ flavour symmetry for the meson and the baryon octect are
written in terms of the meson $\Phi$ and baryon $B$ spinor fields respectively
and are given by
\begin{eqnarray}
	\Phi =  \begin{pmatrix} 
 \frac{1}{\sqrt{2}} \pi^0 +
	\frac{1}{\sqrt{6}} \eta & \pi^+ & K^+  \cr  \pi^- & -
	\frac{1}{\sqrt{2}} \pi^0 + \frac{1}{\sqrt{6}} \eta & K^0  \cr 
	K^- & \bar{K}^0 & - \frac{2}{\sqrt{6}} \eta 
\end{pmatrix}  \, , \quad
	B =
\begin{pmatrix} 
	\frac{1}{\sqrt{2}} \Sigma^0 + \frac{1}{\sqrt{6}} \Lambda &
		\Sigma^+ & p \cr 
		\Sigma^- & - \frac{1}{\sqrt{2}} \Sigma^0 
		+ \frac{1}{\sqrt{6}} \Lambda & n \cr 
		\Xi^- & \Xi^0 & - \frac{2}{\sqrt{6}} \Lambda
\end{pmatrix}  \, .
\end{eqnarray}
respectively. These interpolating fields are not unique as they
contain contributions from off-shell states, but they fulfill free
Klein-Gordon and Dirac equations respectively.}. See e.g.
\cite{Nieves:2001wt,GarciaRecio:2002td} for applications and
\cite{GarciaRecio:2005hy} for extensions to $SU(6)$ in hadronic
reactions.

The Chiral Lagrangian preserves the Baryon current 
\begin{eqnarray}
\partial_\mu {\rm tr } (\bar B \gamma_\mu B) = 0 \, . 
\end{eqnarray}
Therefore, for any scalar and SU(2) 
invariant field ${\cal M}$ we have 
\begin{eqnarray}
{\rm tr } (\bar B \slashchar{ \partial} {\cal M} B) = \partial_\mu \left[
 {\cal M} {\rm tr } (\bar B \gamma_\mu B) 
\right] \, , 
\end{eqnarray}
which yields a vanishing contribution to the action since it is a
total derivative.

\section{Reparameterization invariance}

The unitarity condition $u^\dagger u = 1 $ implies that in general we
may use the polar decomposition, 
\begin{eqnarray} 
u=e^{i H} = 1 + i H +  \cdots \, , 
\end{eqnarray}
 where
$H=H^\dagger$ is a hermitian matrix.  In the case of $SU(2)$ one
usually writes 
\begin{eqnarray} 
H = \vec \tau \cdot \vec \phi \, , 
\end{eqnarray}
with $\vec \tau$ the Pauli matrices which imply 
that ${\rm Tr} H=0$. This standard choice is not unique, 
and in particular one may take 
\begin{eqnarray}
H  = \vec \tau \cdot \vec \phi  + \xi \vec \tau \cdot \vec \phi  (\vec \phi \cdot \vec \phi) + {\cal O} (\phi^5) \, , 
\end{eqnarray}
which complies equally well the unitarity of $u$. The arbitrariness of
$\xi$ will be exploited below.  While the non-linear character of $u
(\phi)$ generally implies the appearence of many body forces, they
turn out to be chirally supressed by the pion weak decay constant. On
the other hand, fields appearing in an EFT are not
unique since one has the freedom to make a change of variables or
field redefinition $\phi \to \phi'=F(\phi) $ with the same quantum
numbers with no consequences on the physical S-matrix. This is the
so-called equivalence theorem.

The lack of reparameterization invariance shows up in final results
only if incomplete calculations are carried out. For instance, the
description of hadronic resonances makes the use of unitarity
mandatory. Any unitarization procedure corresponds to an infinite but
partial sum of Feynman diagrams and the reparameterization invariance
is violated {\it after} unitarization (see e.e.g \cite{Nieves:1999bx}
for the case of $\pi\pi$ scattering within a Bethe-Salpeter framework.).

\section{Cancellation of three body forces in SU(2)}

The problem of existence of three-body forces in the Baryon-Meson
system relies on whether or not there is a particular choice of the
arbirary variable $\xi$ where the terms in the Lagrangian with four
pion fields $\vec \phi$ vanish.  Actually any value of $\xi$
corresponds to a specific perturbative choice of coordinates on the
$SU(2)$ group around the origin. Expanding in powers of $\phi$ we get
in the effective Lagrangian Eq.~(\ref{LB1}) with Eq.~(\ref{LB1_exp})
\begin{eqnarray} 
\Gamma^\mu = \Gamma_2^\mu + \Gamma_4^\mu + {\cal O} (H^6) \, , 
\end{eqnarray}
where
\begin{eqnarray} 
\Gamma_2^\mu  &=& \frac12 \left[H,
  \partial^\mu H \right] \, ,  \\
\Gamma_4^\mu &=& \frac16 H^3 \partial^\mu H + \frac14 H^2
\partial H^2 - \frac16 H \partial^\mu H^3 + \frac1{24} \partial^\mu
H^4 \, . 
\end{eqnarray}
 Now, to the desired order in $SU(2)$ we get (sandwiching with B-fields and partial integration understood)
\begin{eqnarray}
\Gamma_2^\mu &=& i ( {\bf \varphi} \wedge \partial^\mu {\bf \varphi} ) \cdot \tau \left[ 
1 + 2 \xi {\bf \varphi}^2\right] + {\cal O} ({\bf \varphi}^6) \, , \\ 
\Gamma_4^\mu &=& \frac16 {\bf \varphi}^2 \left[ (\tau \cdot {\bf \varphi}) , \partial^\mu
  (\tau \cdot {\bf \varphi}) \right] + \frac16 {\bf \varphi}^2 (\tau \cdot {\bf \varphi})
\partial^\mu (\tau \cdot {\bf \varphi}) + \frac14 {\bf \varphi}^2 \partial^\mu ({\bf \varphi}^2) \nonumber \\ 
&&- \frac16 (\tau \cdot {\bf \varphi}) \partial^\mu \left[ (\tau \cdot {\bf \varphi})
  {\bf \varphi}^2 \right] + {\cal O} ({\bf \varphi}^6) \, . 
\end{eqnarray}
Finally, discarding total derivatives we
get after some algebra 
\begin{eqnarray}
\Gamma^\mu = i ( {\bf \varphi} \wedge \partial^\mu {\bf \varphi} ) \cdot \tau \left[ 
1 + (2 \xi+\frac16 )  {\bf \varphi}^2\right] + {\cal O} ({\bf \varphi}^6) \, .  
\end{eqnarray}
If we choose $\xi=-1/12$ the terms with four pions cancel. Note that
in our case the cancellation does not make use of the equations of
motion and hence holds off-shell. As a consequence, any sub-diagramm
containing these contributions will cancel. Thus, for this field
coordinates there are no three body $\pi\pi N $ forces at order
$1/f^4$ in the chiral Lagrangian.

In a series of
works~\cite{MartinezTorres:2007sr,Khemchandani:2008rk} the study of
three hadron resonances with baryon number $B=1$ was vigorously 
started within a unitary approach based on the Faddeev equations.  The
kind of cancellation found above complies with the result found in
Ref.~\cite{MartinezTorres:2007sr,Khemchandani:2008rk} where a direct
analysis of the Faddeev equation using the standard polar field
coordinates (corresponding to $\xi=0$) and using the on-shell
conditions corresponding to the equations of motion. This
simplification is crucial as it reduces the complicated analysis of
the three body problem to a more feasible linear algebraic value
problem.

\section{Discussion and outlook}

The suppresion of three body forces was a major original motivations
to introduce EFT approaches based on the chiral symmetry of QCD in
Nuclear Physics~\cite{Weinberg:1992yk}. Actually, the possibility of
computing Pion-Deuteron scattering in a model independent leads to the
absence of three-body corrections at threshold at order ${\cal
  O}(1/f^4)$ since ``the sum of all corrections vanishes for a variety
of reasons, among them the threshold kinematics and the isoscalar
character of the deuteron'' after an intricate diagramatic
analysis~\cite{Beane:2002wk} which might be simplified using a
suitable field reparameterizations as done here.

As a final remark we note that {\it local} field redefinitions in EFT
are innocuous in dimensional regularization where the functional Jacobian
vanishes. Unfortunately, the application of this regularization is
subtle beyond perturbation theory. Actually, finite cut-offs may
jeopardize the reparameterization invariance. This is a potential
drawback inherent to the framework, and relevant when implementing
exact unitarity (see e.g.~\cite{Arriola:2010tu} as applied to
ultracold atomic systems and the interplay with van der Waals forces)
which needs clarification.

\section*{Acknowledgements}

I thank A. Mart{\'\i}nez Torres for remarks. This work is
supported by the Spanish Mineco (Grant FIS2014-59386-P) and Junta de
Andaluc{\'\i}a grant FQM225-05.

\end{document}